# Irregular Motion of a Falling Spherical Object Through Non-Newtonian Fluid

Samira Hasani[1], Nahid Maleki-Jirsarei[1] and Shahin Rouhani[2]

[1] *Complex Systems Lab., Physics Department, Faculty of Science, Azzahra University, Tehran, Iran*

[2] *Physics Department, Sharif University of Technology, Tehran, Iran*

## Abstract

The falling of an object through a non-Newtonian fluid is an interesting problem, depending on the details of the rheology of the fluid. In this paper we report on the settling of spherical objects through two non-Newtonian fluids: Laponite and hair Gel. A falling object's behavior in passing through a thixotropic colloidal suspension of synthetic clay, Laponite, has been reported to have many behavioral regimes. Here we report observation of a new regime where irregular motion is observed. We argue that this irregular motion may be interpreted as onset of chaos. Observation of this regime depends on the size of the falling sphere, relaxation time of fluid and concentration of particles in the suspension. Similar experiments in Gel, a yield stress polymeric fluid, do not reveal such behavior.

## 1- INTRODUCTION

The falling of an object through a Newtonian fluid is a classical problem in fluid mechanics. The drag force experienced by a falling ball of radius R, when moving with terminal velocity U, through a fluid with viscosity μ is $F_D = 6\pi\mu\, R\, U$ ,[1]. Though fundamentally correct, this formula has been adjusted through modern considerations [2,3,4,5] and forms the basis of measuring viscosity in a Newtonian fluid. In recent years, the settling of a solid object through a non-Newtonian fluid has received attention [6,7,8], due to various rheological and structural properties of non-Newtonian fluids, the problem of a settling sphere in non-Newtonian fluid is less known. A falling sphere in a non-Newtonian fluid behaves in a complex fashion, in short term it does not always reach a terminal velocity [9], though it may come to a complete rest, sometimes it approaches a terminal velocity, after an oscillating transient [7]. In this paper we present evidence that a sphere falling in Laponite may experience an irregular even chaotic motion. This is in contrast to the settling of a sphere in a yield stress fluid such as Gel (Carbopol) where a simpler behavior is observed.

We have performed experiments in two fluids, Laponite (B) and hair Gel (Carbopol). Hair Gel mainly is composed of Carbopol and PVP (poly vinylpyrrolidone) and it is a polymeric, yield-stress fluid. It exhibits no significant thixotropy or aging in its bulk properties. Laponite is smectite hectorite clay. The Laponite particle has a disk shape typically with diameter of 25nm and thickness of 0.92 nm. There are different grades of Laponite for a complete listing see [10]. We used Laponite B, which is a synthetic layered fluorosilicate and has diameter 40 nm [11]. When Laponite powder is dispersed in water, they swell to form a colloidal dispersion. Laponite phases depend on its concentration and time of rest (aging). Different Laponite phases discussed in literature are characterized by dynamic and static light scattering, small-angle neutron scattering (SANS) and X-ray scattering (SAXS) [12,13,14].

Usually Laponite is studied in two ranges of particle densities, at high particle densities; it has a 'House of Cards' structure. This is based on short- range part of the electrostatic interaction particles [10]. At low particle densities, clusters are formed in the Laponite. There are two different views on clusters, firstly, discussed by Lu et al.[15] that clusters consist of fractal networks and low – density Gel may consist of the network of clusters, secondly discussed by Bonn et al. [16] when they filtered the Laponite suspension ,they don't find evidence for a fractal like organization of the particles. They showed that observation of clusters with fractal network is an artifact, and Laponite behave as a glass. In both scenarios, when time is passed, clusters aggregate and form larger clusters.

## 2- EXPERIMENTAL SET UP

Gel is a polymeric, yield stress, shear thinning fluid ($\rho_g$=1130kg/m$^3$) [17], whereas Laponite is a synthetic clay frequently used as a model aging material which is a thixotropic fluid ($\rho_l$=1109kg/m$^3$) [18]. Figure 1 shows shear stress vs. shear rate and figure 2 shows viscosity vs. shear rate for Laponite and gel.

The Laponite powder is mixed slowly with swirling distilled water approximately at PH 8 for 1 hour with mechanical stirrer [Heidolph RZR 2041] at 500 rpm. For 10 days, the fluid is stirred every 2 days for 1 hour at 500 rpm. Just before the beginning of the experiment the fluid is mixed (1 hour at 500 rpm) in order to obtain reproducible initial state.

We used Laponite suspension with concentration 1.5wt %, and experiments were repeated with different resting times. This concentration is the lowest concentration for which Laponite stays a Gel-like for the resting times we have experimented with, lower concentrations one may run into liquid regime [19]. Selected steel sphere did not move in Laponite of concentration 3wt%.

Hair Gel is composed of Carbopol, PVP, uvinul and preservative it was mixed and prepared with water and for 1 hour was swirled 500 rpm at 100 C. Later it was swirled for 3hours, without any heating until its bubble go out.

Prepared fluids were put into a cylindrical vessel with a depth of 45 cm and an inner diameter of 6 cm (D), the fluid depth is 40 cm into the cylinder. Laponite was left at rest a given time during which it could restructure so that its apparent viscosity increased. Then a steel sphere was put with pliers in the fluid just below the free surface and closes to the central axis of the vessel and let go. The settling sphere process is recorded with a CCD camera. All motions are observed at low Reynolds number ( $Re < 10^3$ ), where flow is laminar [1]. We used steel spheres in different sizes ($2.5mm < d < 14mm$) and same density ($\rho_s$=7753kg/m$^3$). The ratio of sphere to vessel radius ($0.04 < \frac{d}{D} < 0.2$) is small, so wall effects are negligible [20].

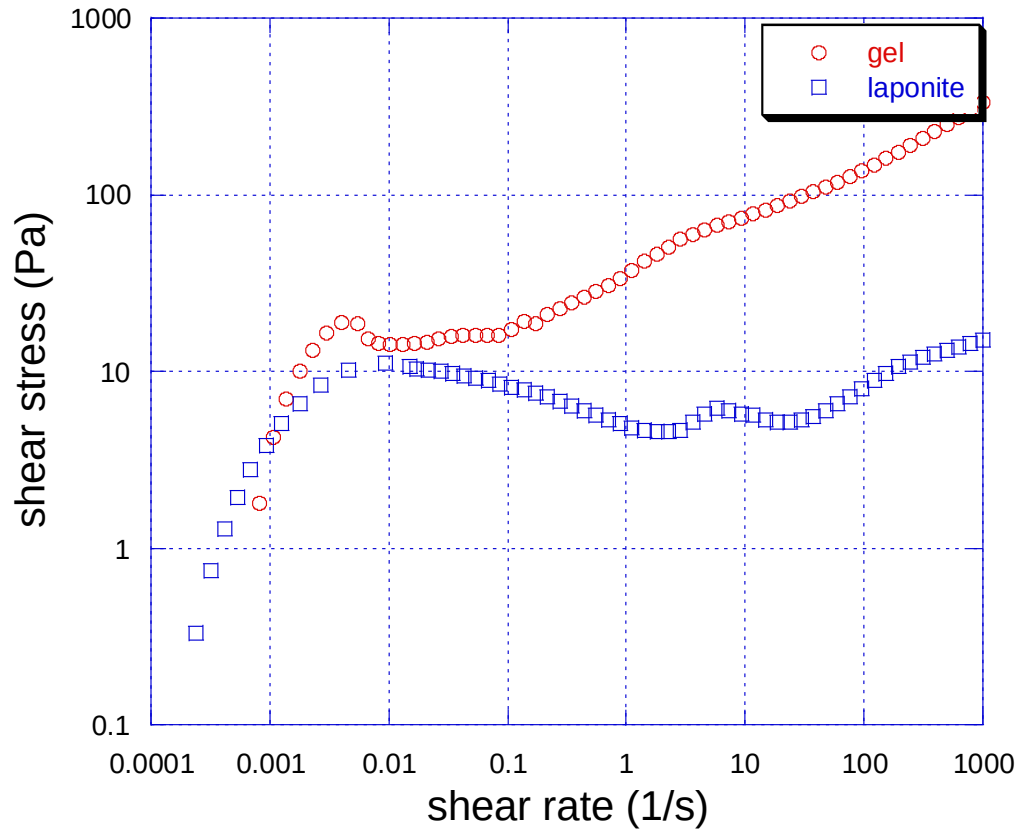

FIG. 1. Shear stress vs. shear rate for Gel and Laponite (1.5wt %)

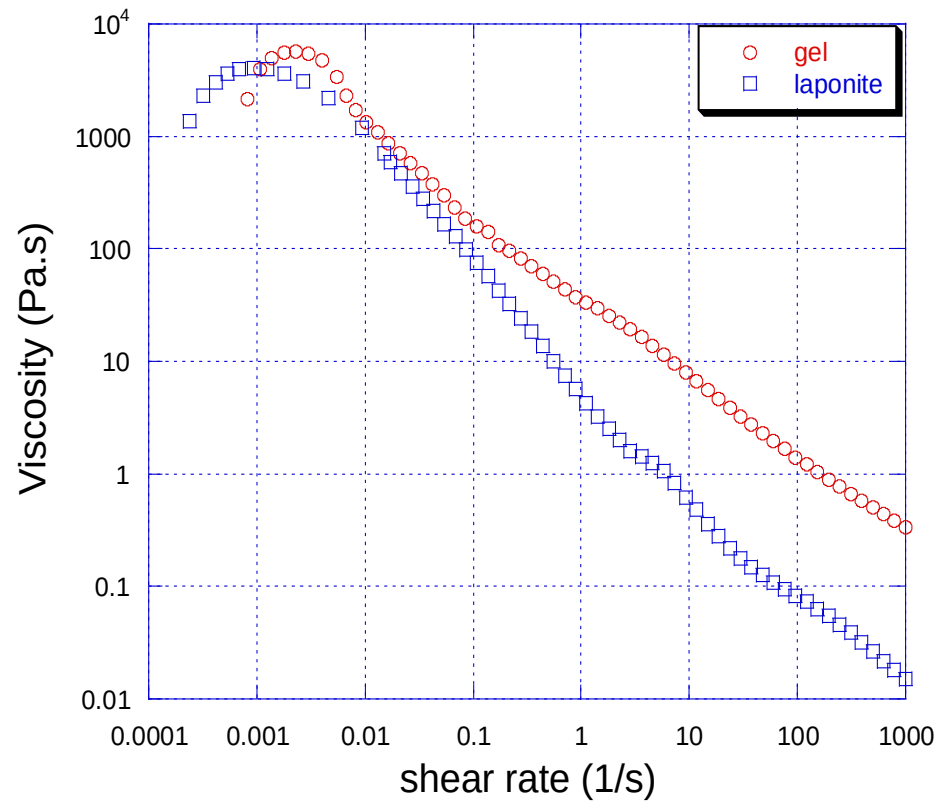

FIG. 2. Viscosity vs. shear rate for Gel and Laponite (1.5wt %)

## 3- OBSERVATIONS

Observing the settling of steel balls (diameters ranging from 4 to 14 mm) in polymeric Gel, we observed a terminal velocity, though some oscillations are observed for small radii. Figures 3 and 4 show distance of sphere measured from free surface of fluid, vs. time. The terminal velocity is the slope of the curve at long times. These oscillations were observed before and they are believed to be due to interaction between the inertia of the sphere and the elastic behavior of the material in its solid regime [3].

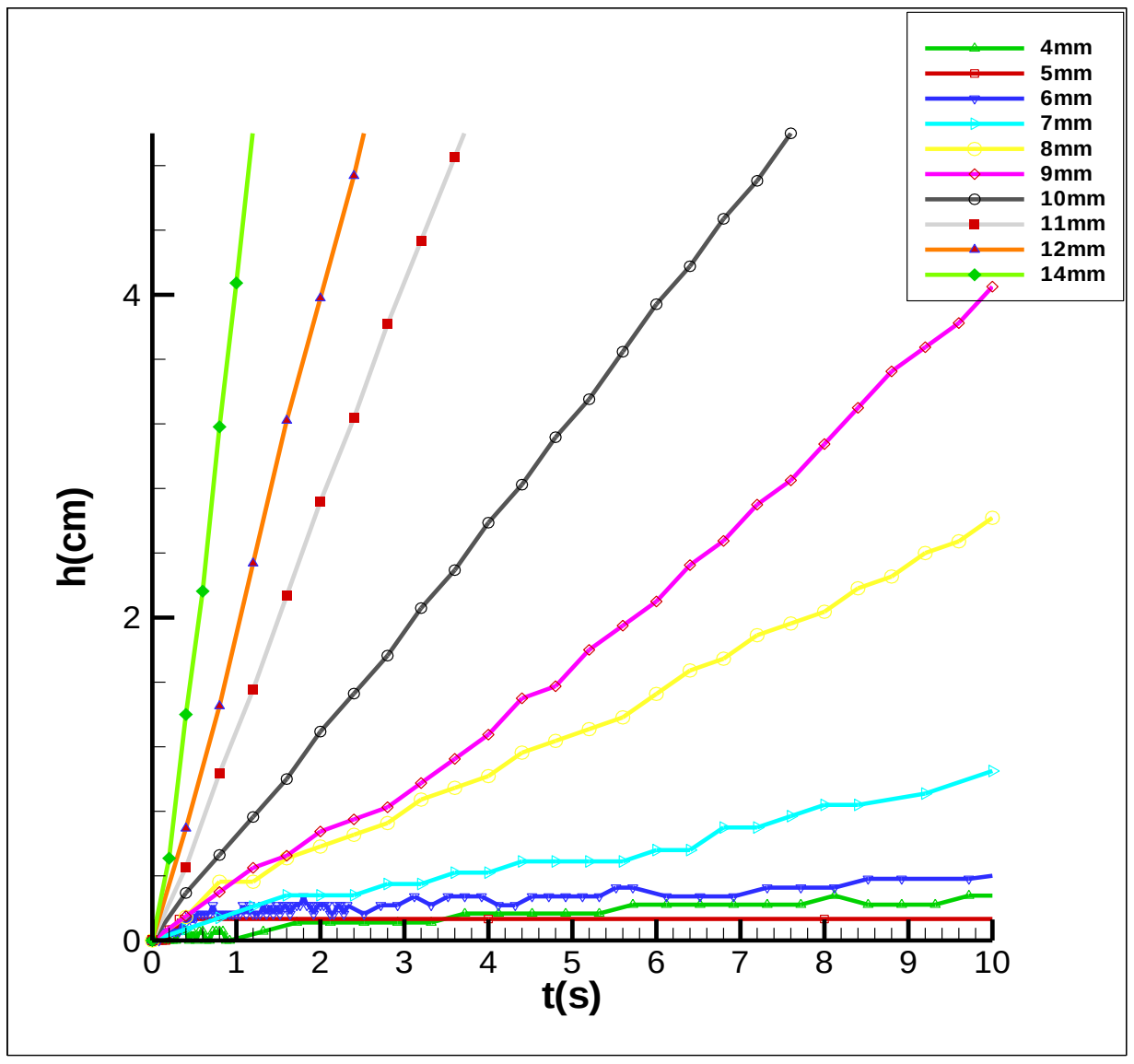

FIG. 3**.** Distance versus time for falling spheres through polymeric Gel starting from free surface of fluid.

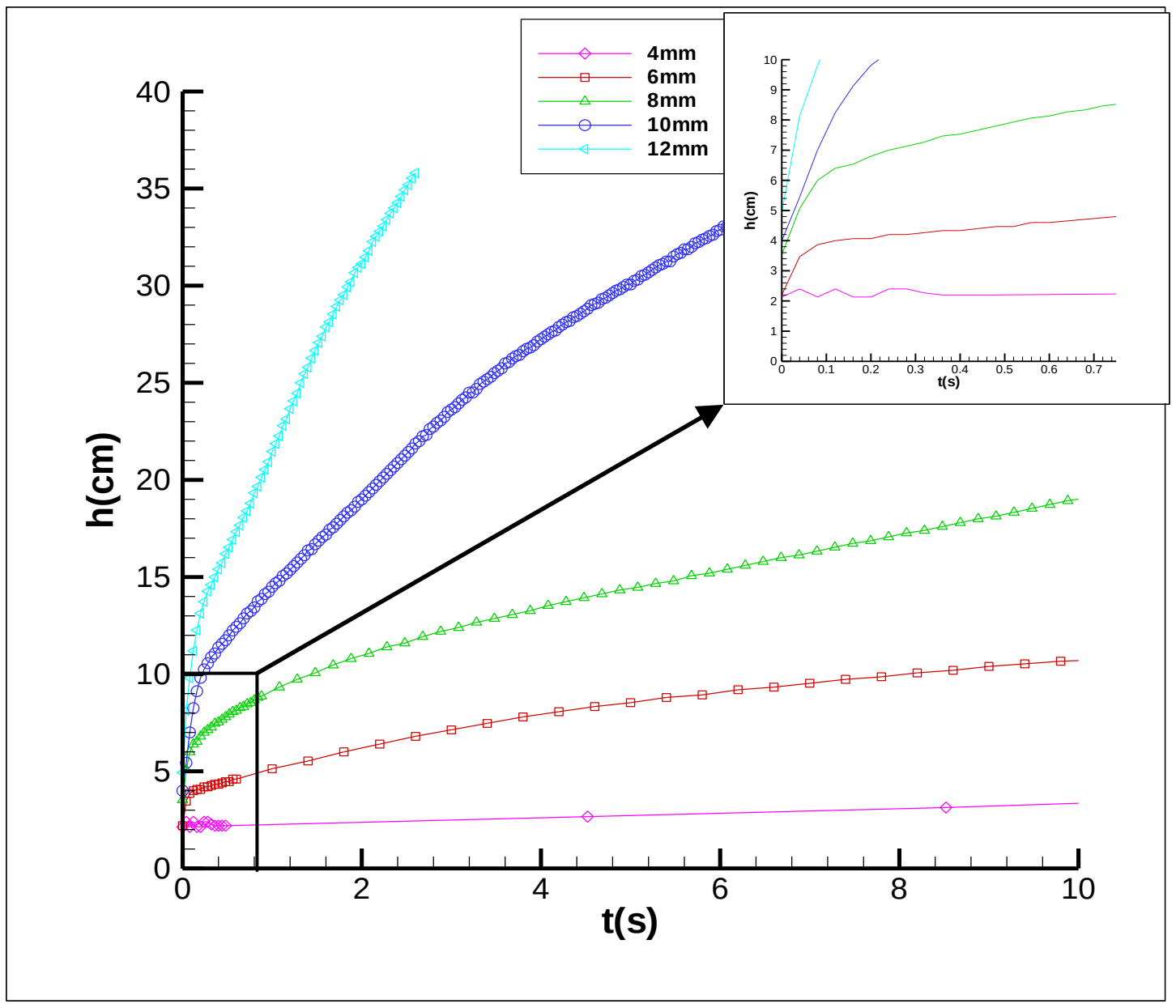

FIG. 4. Distance versus time for falling spheres through polymeric Gel when balls are dropped about 5 cm above the free surface. (Inset shows same date in larger scale)

Spheres with diameters less than 4mm don't move into this polymeric Gel or they move very slowly and stop. Spheres with diameter bigger than 4mm mostly move with constant velocity. For spheres that their diameter is bigger than 5 mm, oscillations aren't observed and they move always with large velocity.

For a sphere object settles through the Laponite suspension, literatures explained three regimes for its motion.  For short times of rest, settling spheres move through the Laponite always at constant, large velocity (regime 1). For intermediate times of rest, spheres move at intermediate velocity which continuously decreases (regime 2). And for long times of rest, they move at very low velocity (regime 3) [21].We observed in our experiments, other than three regimes that are explained above, there is another regime for settling object through the Laponite. When small steel spheres (their diameters are 3mm and 4 mm) are selected and experiments are done at various times of rest ($t_w$), settling sphere through the Laponite does not always move at a constant velocity or acceleration. There are oscillations at velocity of settling object in different times of rest (figure 5). Time of rest and size of balls are important factors. Also small object falling through Laponite does not move on a straight line to reach the vessel bottom [22].

Additional experiments were done on various diameters of steel spheres: 1.5, 2, 2.5, 3, 3.5,4 and 4.5 mm. For small spheres with diameter less than 2.5 mm, balls move slowly or they don't move through the fluid at all (regime 3). Spheres with diameters between 2.5 and 3.5, balls do not move with constant velocity and have oscillatory motion. Spheres with diameters bigger than 3.5mm move with large constant velocity through the fluid (regime 1). These results were achieved for small time of rest (10 minutes). Figure 5 shows experiments done in different times of rest.

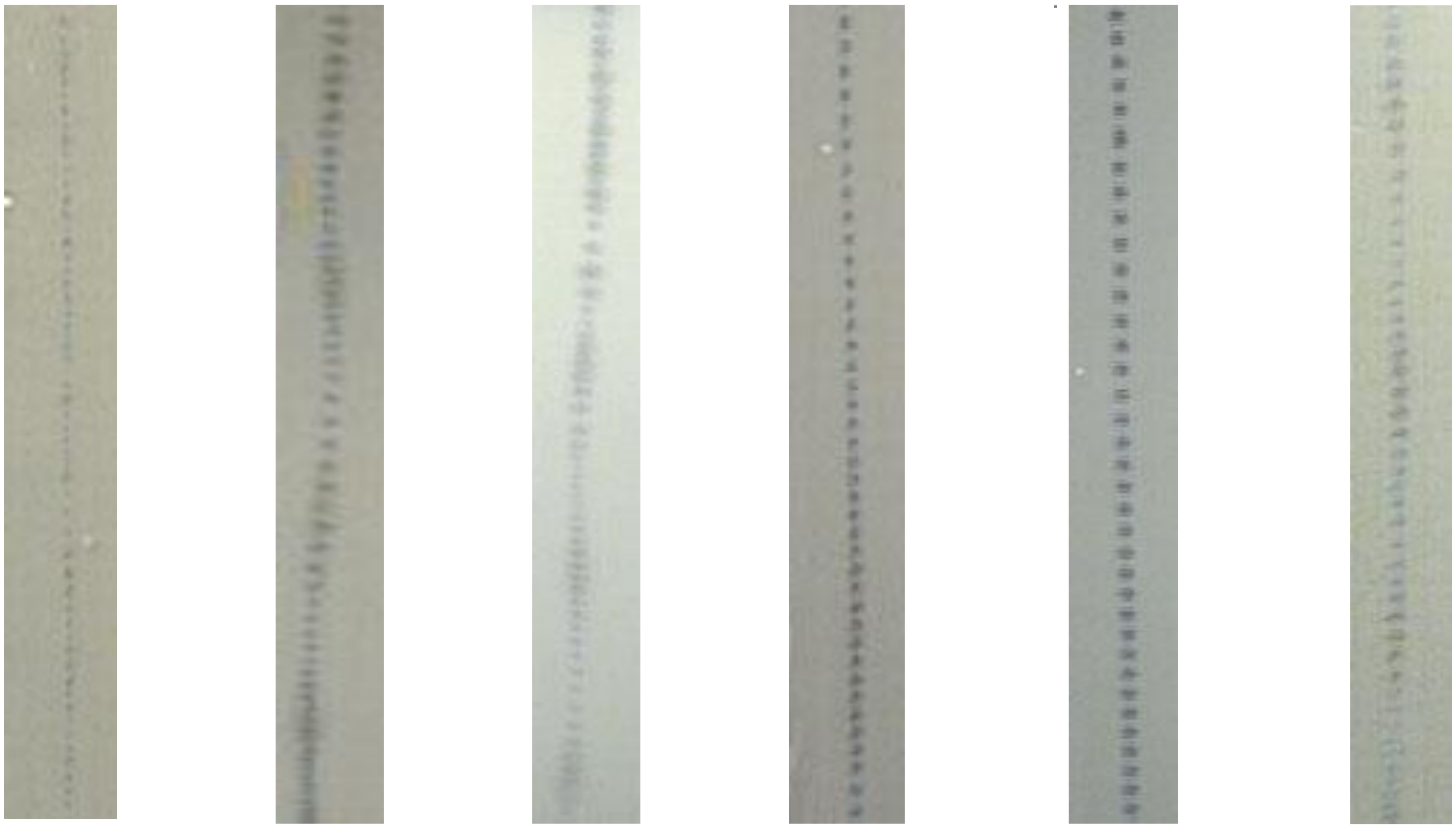

(a) (b) (c) (d) (e) (f)

FIG. 5. Set of video images of steel sphere falling through Laponite 1.5% wt. For the sake of distinguishable images time interval is different for each image. (a) 2.5mm diameter steel sphere at $t_w$ = 10 minute and time interval is 8s. (b) 3mm diameter steel sphere at $t_w$ = 10 minute and time interval is 0.4s. (c) 3mm diameter steel sphere at $t_w$ = 30 minute and time interval is 1.6s. (d) 3.5mm diameter steel sphere at $t_w$ = 10 minute and time interval is 0.2s.(e) 4mm diameter steel sphere at $t_w$ = 10minute, and time interval is 0.12s. (f) 4mm diameter steel sphere at $t_w$ = 30 minute and time interval is 0.4s

More experiments were done between two glass walls. There wall effect is too large but small falling spheres don't move with constant velocity.

## 4- ANALYSIS

Distance of settling sphere from free surface and its velocity is computed using motion analyzer and commercially available software. We noted that the motion in Gel was fundamentally different from Laponite. Graphs for velocity of sphere falling through the Laponite suspension as a function of time shows these differences. Figures 6-9 show irregular motion of a falling spherical object through Laponite suspension.

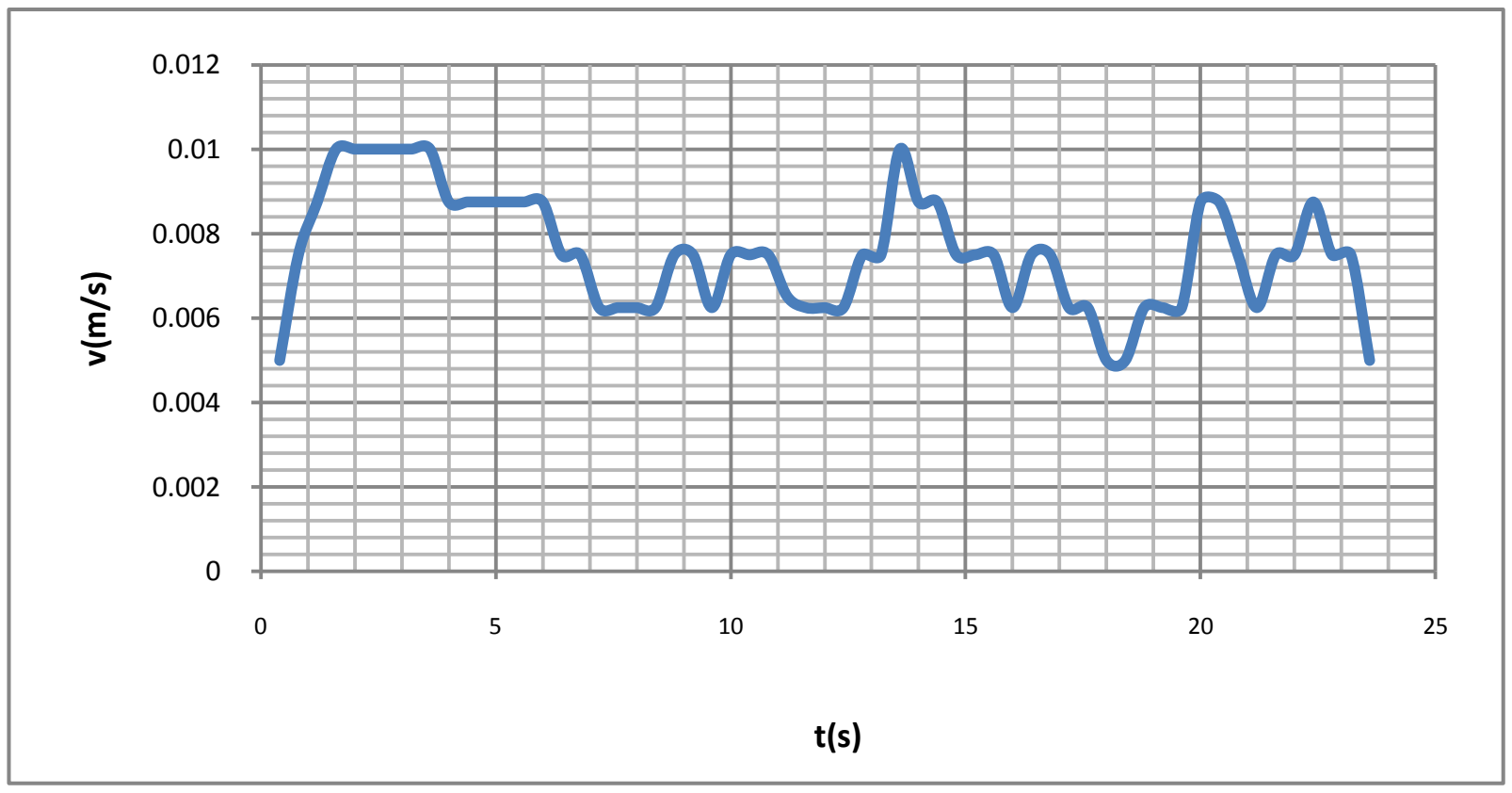

FIG. 6. Velocity as a function of time for 3mm diameter sphere falling through the Laponite ($t_w$ =10 min)

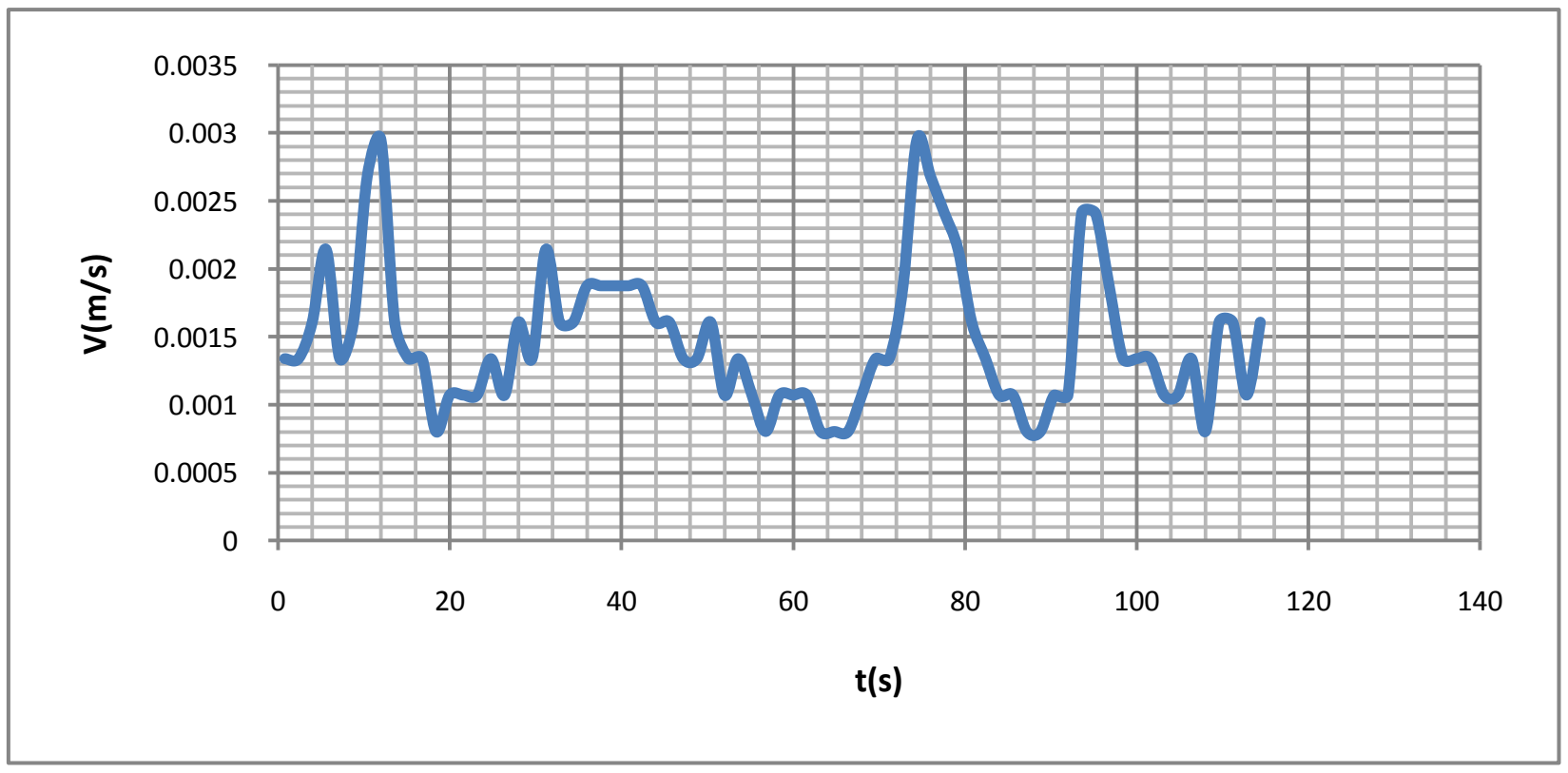

FIG. 7. Velocity as a function of time for 3mm diameter sphere falling through the Laponite ($t_w$=30 min)

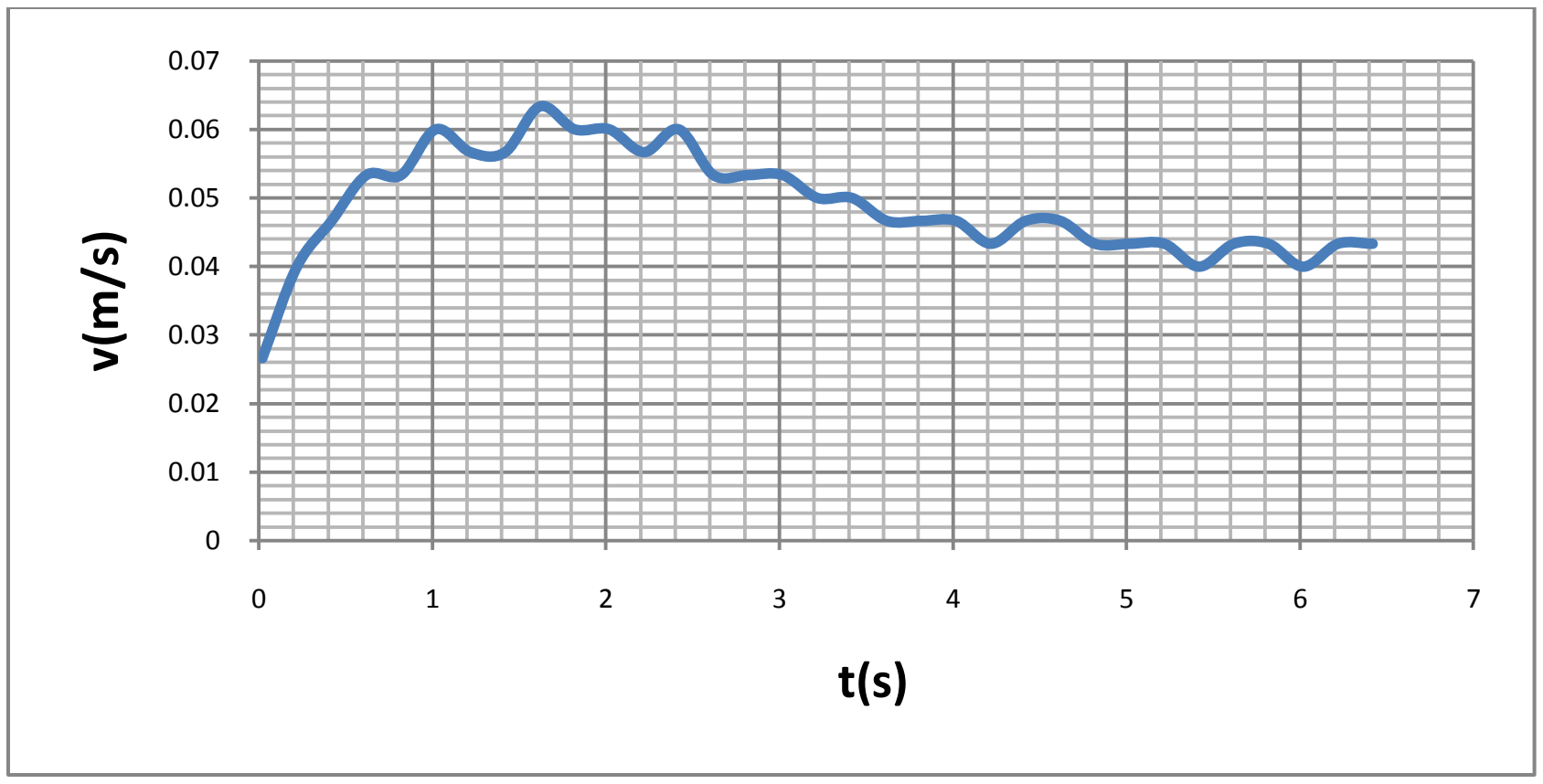

FIG. 8. Velocity as a function of time for 4 mm diameter sphere falling through the Laponite ($t_w$ =10 min)

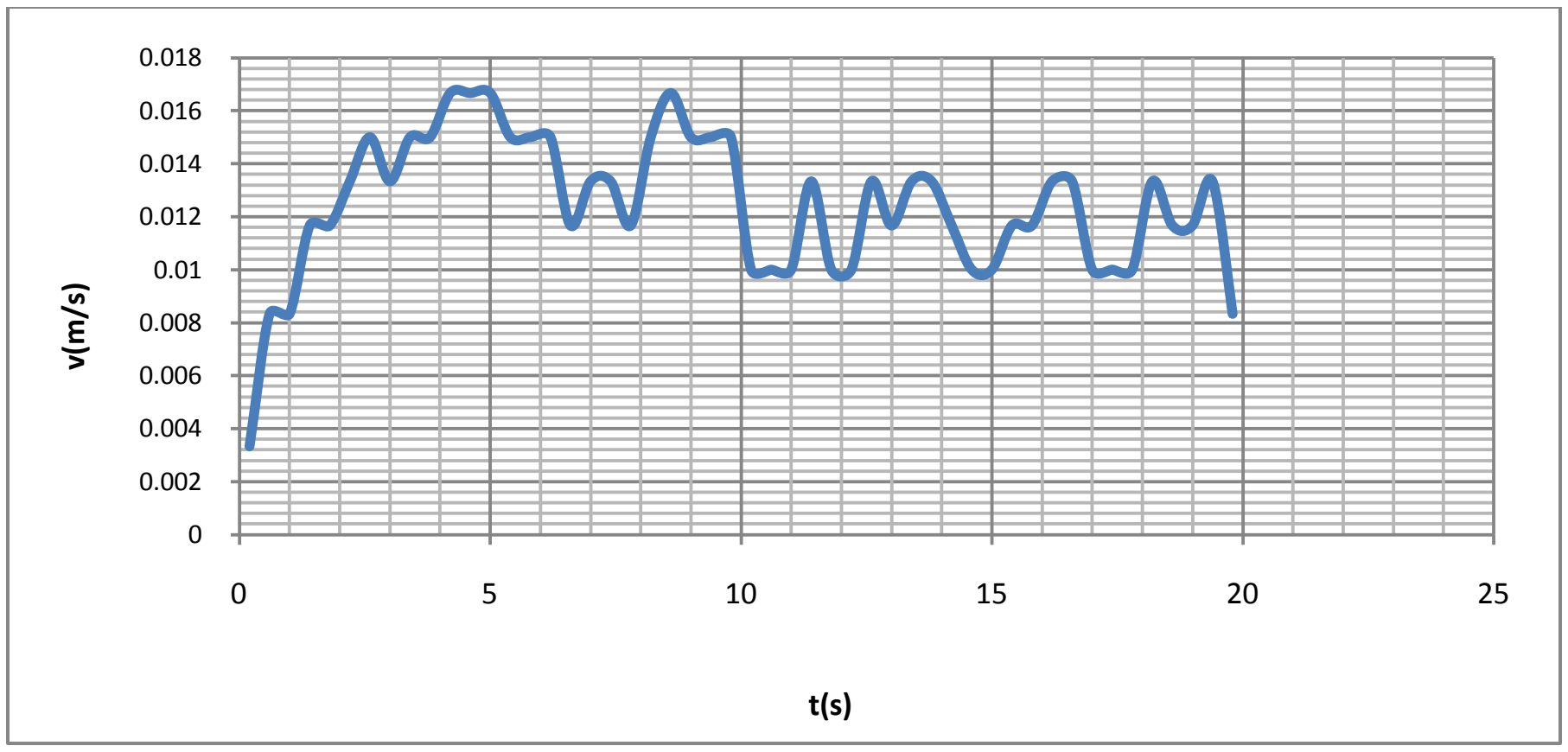

FIG. 9. Velocity as a function of time for 4 mm diameter sphere falling through the Laponite ($t_w$ =30 min)

To describe the motion of the falling sphere through Laponite we appeal to the “House of Cards” structure, perhaps the ball is falling in stages, rather like a fruit through a tree. Let us assume that the velocity of the falling object is $v_i$ at the end of a suitable interval. Velocity at the end of the next interval $V_{i+1}$ is related to the previous interval is related to the velocity of the previous interval and the innate properties of the fluid. Previous work shows that exert forces on sphere falling through non-Newtonian fluid may be functions of $v_i^2$ and $v_i$ [2-8]. So we consider iterative processes of the form

$$v_{i+1} = f(v_i) = av_i^2 + bv_i + c$$

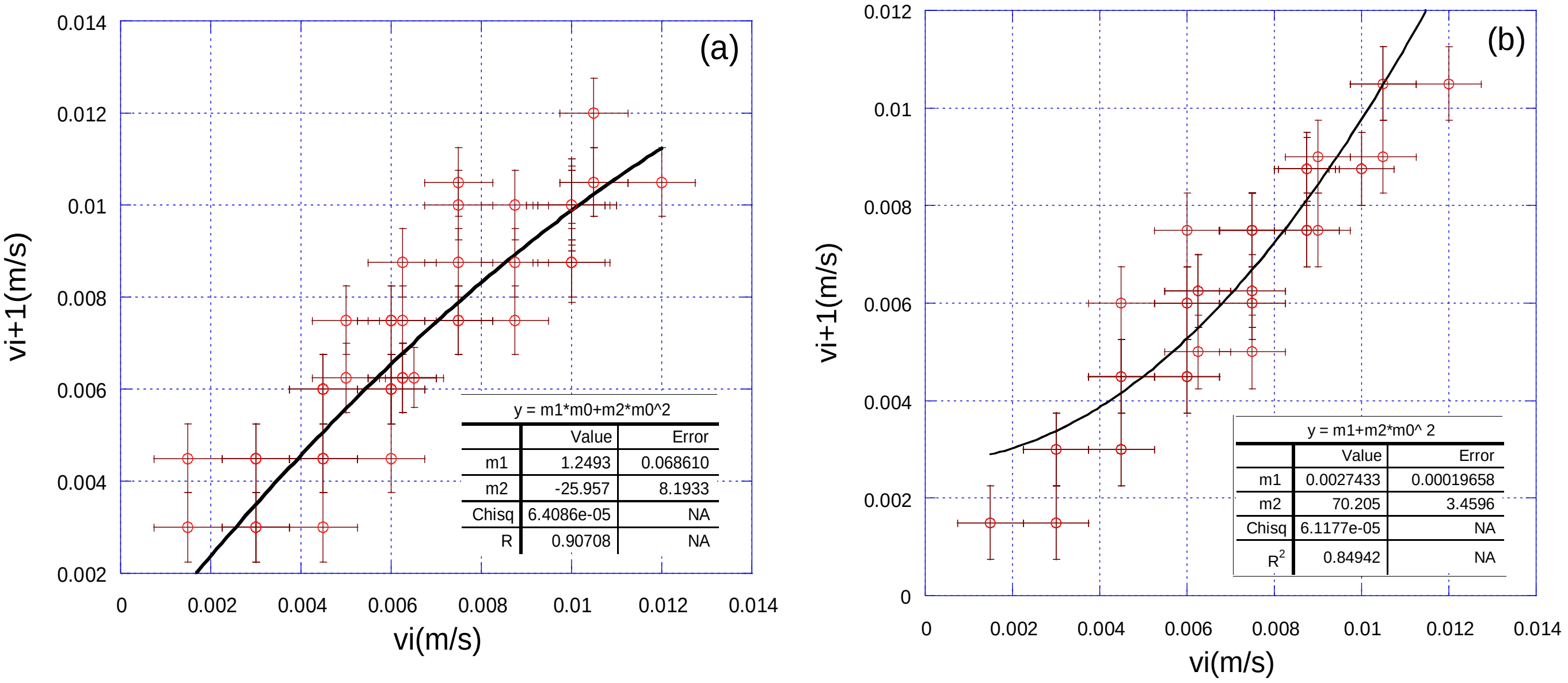

FIG. 10. Diagram displays$Vi_{+1}$ (vertical axes) versus $V_i$ (horizontal axes) for small sphere (diameter of steel sphere is 3mm) falling through the Laponite suspension ($t_w$ =10min) on one cycle. Any cycle divide to two districts. Graph (a) shows when velocity is increasing to its maximum value on one cycle and graph (b) shows when velocity is decreasing to its minimum value at same cycle. Fitting functions in both graphs are function of $v_i^2$and$v_i$ .

Here we have assumed a quadratic $f(v_i)$ whereas more complex functions may be taken. Figure 10 shows $v_{i+1}$ versus $v_i$ for small sphere on one cycle and also shows fitting functions that maintain above model. Any cycle divide to two districts; In Fig. 10 (a) speed increases from minimum to maximum value in one cycle. In Fig. 10 (b) Speed decreases from maximum to minimum on same cycle. Fitting functions show the drag force obtained from these diagrams is comparable with models previously presented [21]. The difference between the two cases is due to the structure of Laponite. Difference between coefficients of square velocity in fitting functions of above cases shows differences in drag forces. Drag force is related to flow induced structure. When this structures breaks, sphere suddenly accelerates [9].

So Figure 11 shows three types of attractors for different spheres falling through a Laponite suspension.

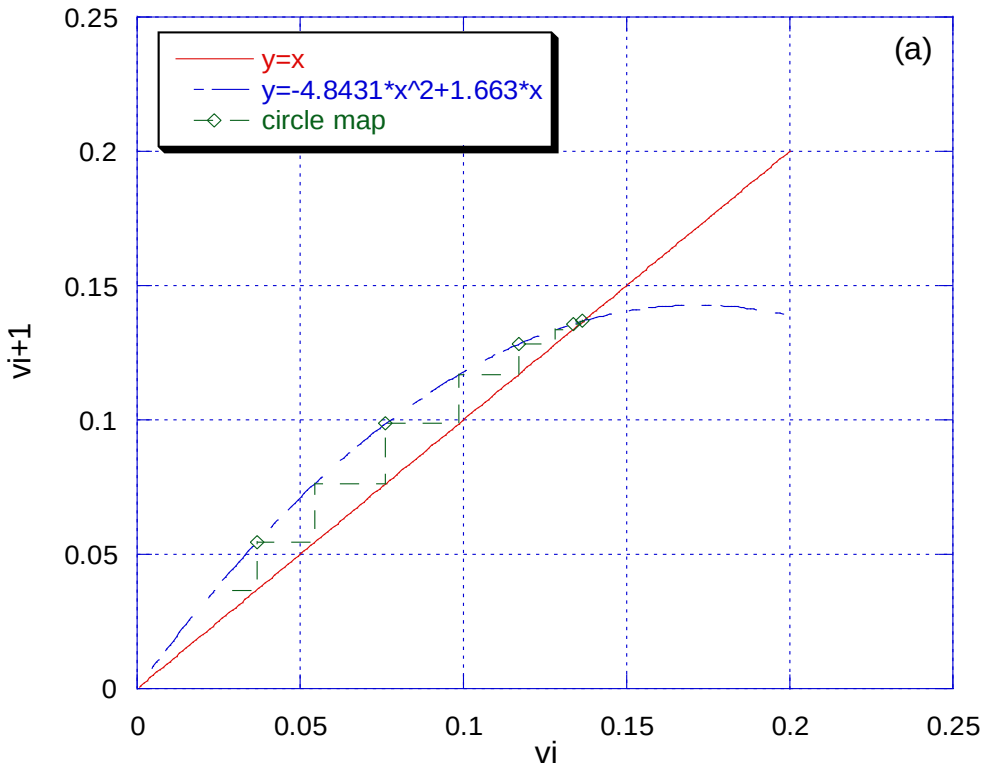

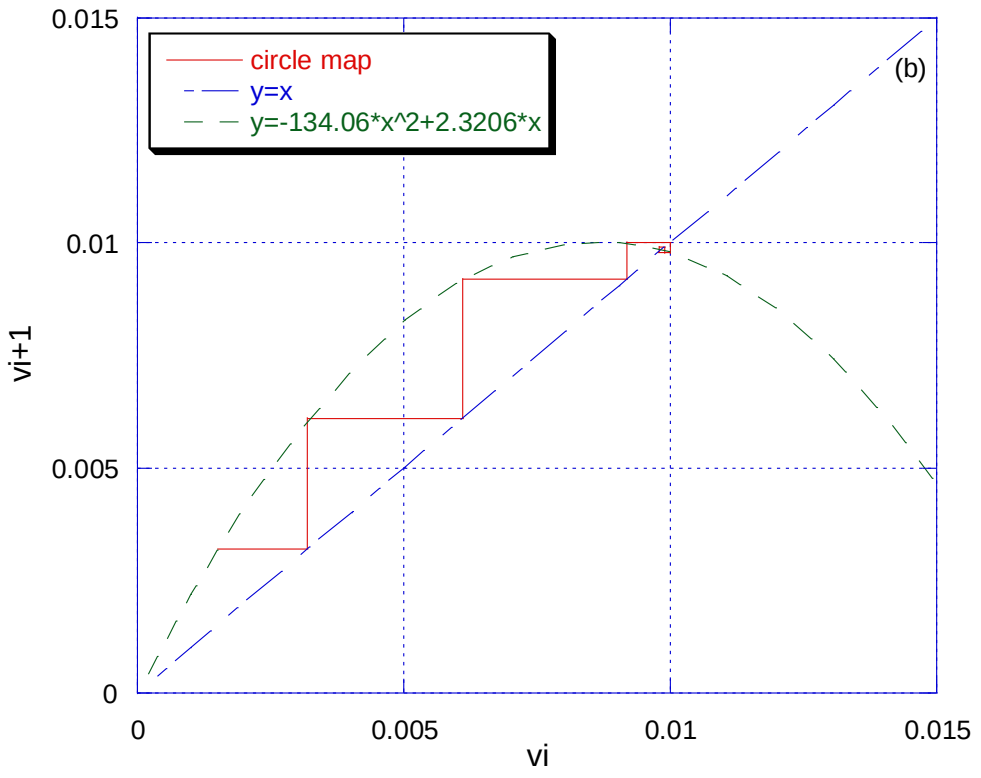

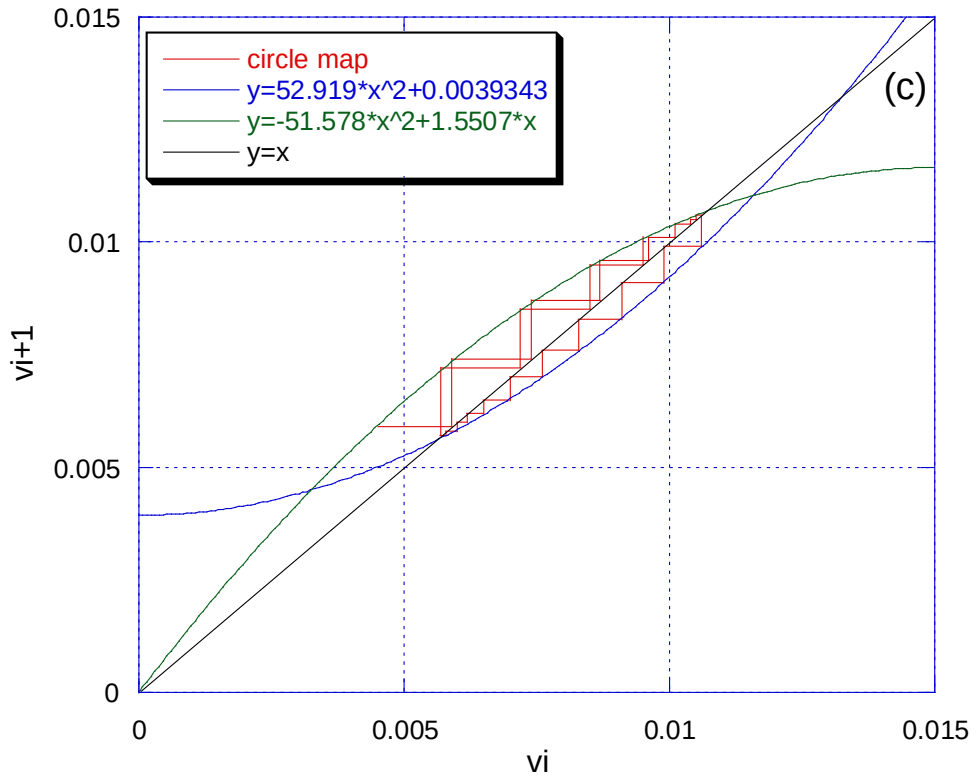

FIG. 11 shows three different behavior of sphere through Laponite suspension a)This graph shows the simple fixed-point attractive behavior for falling sphere with D=4mm and aged time of laponite is $t_w$=10min. b) this graph shows the orbit settles into a two-cycle for falling sphere with D=4mm and aged time of laponite is $t_w$=30min. c) this graph shows the orbit settles into a 26-cycle for falling sphere with D=3mm and aged time of laponite is $t_w$=10min

## 5- TIME SCALE FROM VISCOSITY TO MODULES

The other point is the relative timescale of velocity fluctuations and the relaxation time. The Deborah number De, is the ratio of relaxation time of the fluid ($\tau$) to time scale of process ($t_a$). This dimensionless number is significant for visco-elastic fluids. A large value of De (>1) denotes a dominantly elastic response and De < 1 indicates viscous behavior [24]. We believe that for any cycle of falling and rising sphere velocity, transition between viscous and elastic behaviors happen. And transition happens when De =1 therefore $\tau = t_a$ . Figure 12 shows loss and storage modules and relaxation time versus elapsed time for prepared Laponite suspension. We have used an oscillatory experiment (amplitude oscillatory($\gamma_0$)=1% and frequency 0.1 Hz) to estimate the both modulus. And relaxation time are calculated by $\tau = G'/\omega G''$ [25] where $G'$ , $G''$ are storage and loss modulus respectively. Relaxation time increases with time and it obeys a power law $\tau \sim t_w^{0.2}$ , so when $t_a \sim t_w^{0.2}$ , transition happens. Transition between viscous and elastic behavior is more substantial when small sphere (D=3mm) falls through aged Laponite suspension ($t_w$ =10 min). Comparison between time scale process that obtains from both experiments (above oscillatory Rheology measurement and free fall of spheres through Laponite suspension) shows that they have same order.

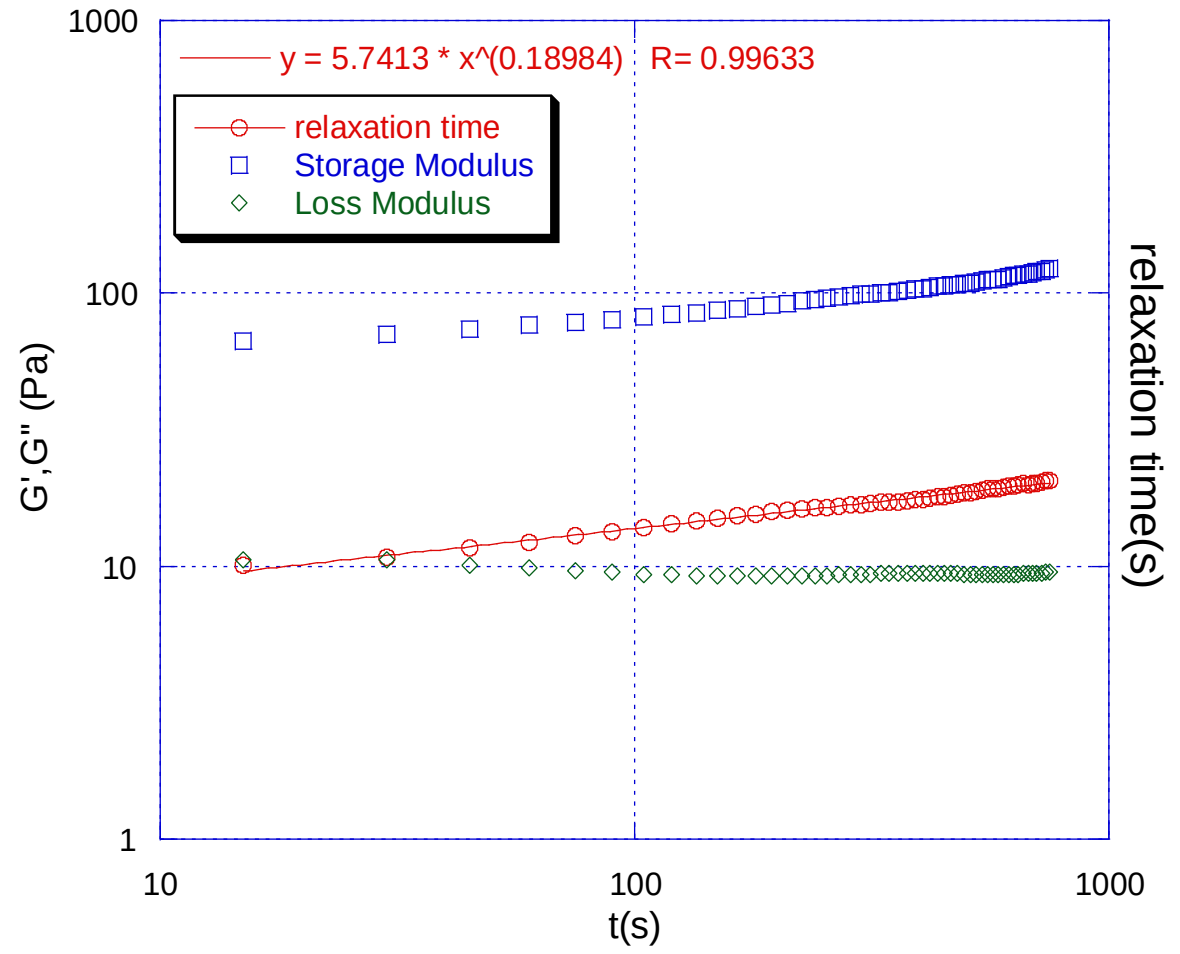

FIG.12. shows storage modulus (G′), loss modulus (G″) and relaxation time as a function of time (frequency= 0.1 Hz, $\gamma_0$=1%).thick red line represents a power low fit to relaxation time – time data ($\tau \sim t_w^{0.2}$).

## 6- CONCLUDING REMARKS

We have presented experimentally various behaviors objects falling through two Non-Newtonian fluids, Laponite and Polymeric gel. Experiments of falling objects through yield stress, polymeric Gel obtain that small objects have oscillations just at the beginning of settling into gel, subsequently in most cases an approach to terminal velocity is made, Figs. 3,4. But motion through Laponite is more complex. Experiments of falling object through colloidal suspension of synthetic clay Laponite show four different regimes. Three of these regimes have been discussed before, in this paper we present a fourth regime showing irregular perhaps chaotic motion. The onset of this regime mostly depends on the size of the falling object and aging of Laponite suspension and concentration of particles. We believe that for any cycle of falling and rising sphere speed, transition between viscous and elastic behavior happens. Transition happens when De ~ 1, so time scale of these speed fluctuations and relaxation time of Laponite suspension are comparable.

## 7- ACKNOWLEDGMENTS

We would like to thank Daniel Bonn for valuable opinions and suggestions on this manuscript. We are grateful to the members of Complex System Laboratory, Azzahra university and financial support from the Azzahra university.

## 8- REFERENCES